# Status of the JET LIDAR Thomson Scattering diagnostic

M Maslov[a], MNA Beurskens[a], M Kempenaars[a], J Flanagan[a] and JET EFDA contributors[*]

JET-EFDA, Culham Science Centre, Abingdon, OX14 3DB, UK

[a] *EURATOM-CCFE Fusion Association, Culham Science Centre, Abingdon, Oxon, OX14 3DB, UK*
[*] *See the appendix of F. Romanelli et al., Proc. 24th IAEA Fusion Energy Conferenc, San Diego, USA, 2012*

E-mail: mikhail.maslov@ccfe.ac.uk

ABSTRACT: The LIDAR Thomson scattering concept was proposed in 1983 and then implemented for the first time on the JET tokamak in 1987. A number of modifications were performed and published in 1995, but since then no major changes were made for almost 15 years. In 2010 a refurbishment of the diagnostic was started, with as main goals to improve its performance and to test the potential of new detectors which are considered as candidates for ITER. During the subsequent years a wide range of activities was performed aimed at increasing the diagnostic's light throughput, improvement of signal to noise ratio and amendment of the calibration procedures. Previously used MA-2 detectors were replaced by fast GaAsP detectors with much higher average QE. After all the changes were implemented, a significant improvement of the measured data was achieved. Statistical errors of measured temperature and density were reduced by a factor of 2 or more, depending on plasma conditions, and comfortably surpassed the values requested for ITER Core Thomson Scattering (10% for $T_e$ and 5% for $n_e$). Excellent agreement with other diagnostics (conventional High Resolution Thomson Scattering, ECE, Reflectometer) was achieved over a wide range of plasma conditions. It was demonstrated that together with long term reliability and modest access port requirements, LIDAR can provide measurements of a quality similar to a conventional imaging Thomson Scattering instrument.

# Contents

## 1. Introduction

LIDAR Thomson Scattering was implemented successfully on JET more than 25 years ago [1,2] demonstrating the advantages of the LIDAR concept: 180 degrees scattering requires only one access port to the plasma, light collection optics do not have to be a high resolution imaging optics, do not require precision alignment and is well suited for a hazardous environment where routine access might be difficult. The spatial resolution is limited by the laser pulse length and detector response time and can be optimized to ~7cm which leads to a spatial resolution of >8% of the minor radius for JET. For larger devices such as ITER this scales to >4% of the minor radius without the need for increasing the number of detectors. Because of these particular advantages, LIDAR was initially chosen as a profile diagnostic on JET and is considered as a candidate for the core plasma Thomson scattering diagnostic on ITER.

The latest publication on developments of the core LIDAR diagnostic on JET was released in 1995[3]. Since then and until very recently, the diagnostic was running without any significant modifications. Recognizing that the data quality produced is far behind of the data produced by other, more modern, diagnostics (e.g. radiometer ECE, conventional high resolution TS) an extensive refurbishment was initiated in 2010 with the main goal to improve the quality of the measurements.

In this paper a description of the LIDAR diagnostic on JET will be given, with special attention to details which represent the key differences between LIDAR and conventional TS design and which were not adequately highlighted in previous publications. Experience in the use of the fast GaAsP detectors and example measurements of the improved performance and a comparison with other JET diagnostics will be given.

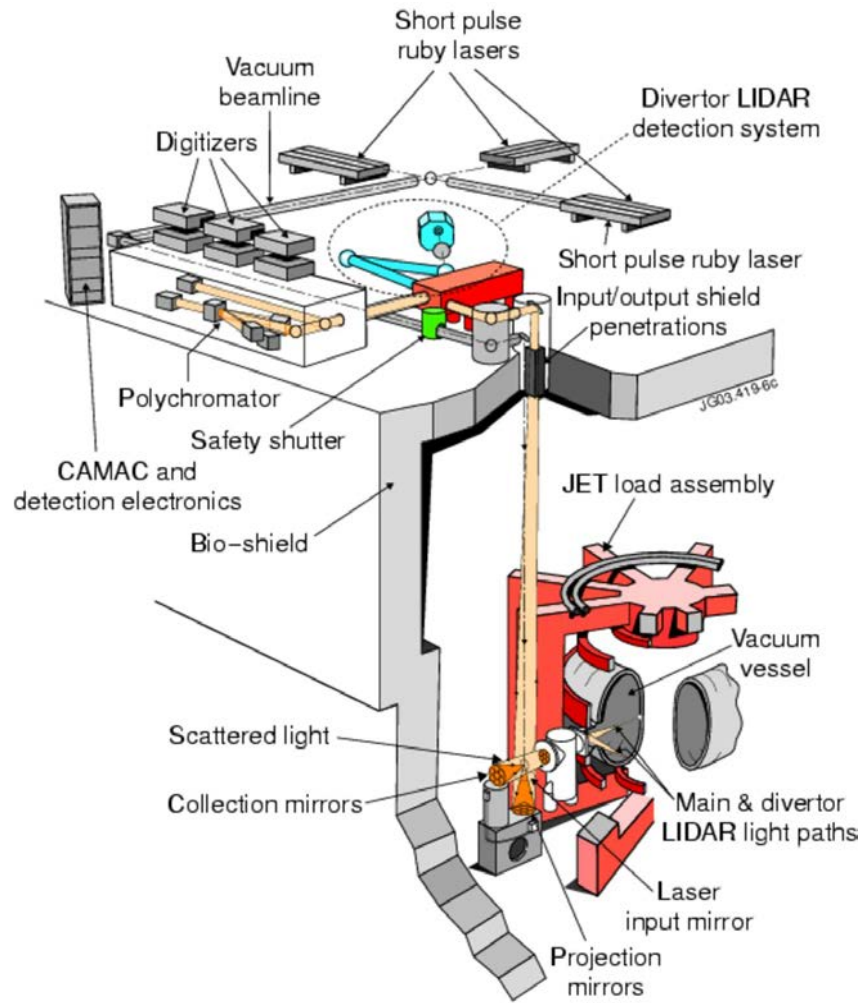
Figure 1: schematic of the JET LIDAR diagnostic

## 2. Scattering geometry and collection optics lay-out

### 2.1 Optics near tokamak

The layout of the LIDAR Thomson Scattering at JET is shown in figure 1. As one can see, all the major components of the diagnostic are located outside of the biological shield (2.5-3m of concrete), whilst only few light collection mirrors are placed near the tokamak. A short laser pulse is generated by a mode-locked ruby laser (694.3nm, ~300ps, ~1 Joule per pulse, 4Hz) and launched via an 8cm wide penetration tube, through the 2.5m biological shield ceiling, towards a 45° mirror in the Torus Hall almost 20 metres below, and then horizontally directed into the JET plasma. Light scattered by the plasma is collected via 6 different vacuum windows with a diameter of d=0.16m located at the end of the JET pumping chamber ~4.4m away from the plasma core. For each of the windows, a separate spherical mirror collects the scattered light and focuses into a small Newtonian mirror, which reflects the light down towards another 6 spherical mirrors which in turn focus the light into another biological shield penetration with a diameter of d=0.25m.

Figure 2 shows the calculated solid angle as if the scattered light was vignetted solely by the vacuum windows or by the collection mirrors. As one can see, in any plasma location the vacuum windows should be the stop aperture. Nonetheless, the real light throughput as measured during Raman scattering calibration is significantly more limited. This strong vignetting is imposed by the narrow penetration through the concrete ceiling of the JET biological shield, thus limiting the radial extent of JET LIDAR profiles. The best light throughput point position can be controlled by adjusting the vertical collection mirrors' radial position, which is currently is set to ~3.5m in order to confidently cover the Low Field side part of the JET plasma ~3.0-3.8m. The High Field side of the profiles is measured with much lower signal to noise ratio, and data for the innermost part may be missing.

The collection optics consist of 6 independent light collection assemblies: 2 spherical mirrors in each and all sharing a single small plane mirror, together forming a Newtonian telescope. In order to maintain alignment stability, all the mirrors are mounted on a concrete tower which is separate from the JET, insuring that any movement of the machine does not affect the optics. Approximately once a year the whole assembly has to be removed to allow remote handling access to the interior of JET vessel and then put back, but no re-alignment of mirrors is normally required after the reposition. That way collection optics can function many years without any adjustments, which is one of the main advantages of the LIDAR concept.

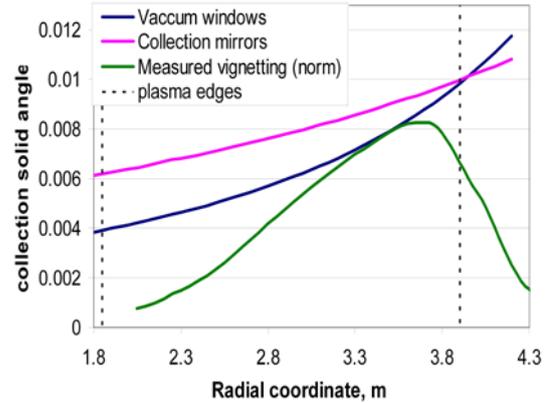

Figure 2: light collection solid angle, seen by vacuum windows and collection mirrors, in comparison with real measured light throughput.

**2.2 Implications on stray light.**

Due to the extremely low Thomson scattering cross section (~$0.7*10^{-24}cm^2$), every single TS instrument has to deal with a combination of very low detectable signal on a background of very bright diffused laser light. Dealing with LIDAR stray light is somewhat different from the conventional TS case. The fast detection system can discriminate various reflections of the short powerful laser pulse as a series of temporally separated spikes. The strongest splash of light is observed as the laser hits the inner wall of the vacuum vessel. Since the impact point of the laser is in direct line of sight of the powerful light collection optics, this so-called "backwall pulse" is traveling straight into the spectrometer and even has the potential to damage sensitive detectors or acquisition electronics. In the previous setup with MA-2 detectors, an overvoltage protection circuit was used. Presently, GaAsP detectors are connected directly to the digitizers (Tektronix TVS645) and no damage has developed so far.

The powerful backwall spike is coming *after* the useful signal and can be easily separated from it, as long as the probing laser does not produce pre-pulses. Pre-pulse rejection of the laser implemented on JET is of the order of $10^{-6}$ and in combination with the spectrometer's stray light rejection is sufficient to reduce their level to not be measurable by the detection system.

The second highest spike is detected at the time when the laser pulse is going through the input vacuum window, this happens before the main signal and would have saturated previously used multi-alkali MA-2 detectors, therefore fast gating of these detectors was necessary. The large distance between the vacuum windows and the JET plasma (~3.5m) serves as a delay between this stray light spike and the light scattered from plasma. This delay is large enough to switch the detectors on and measure the passive background light before the laser pulse reaches plasma. Currently used GaAsP detectors are also gated in the same way, although no tests were performed to estimate the impact of that stray light on the detectors' linearity and it is unknown if such gating is still compulsory.

In the proposed ITER LIDAR design [4], the backwall and the input aperture are the only two sources of stray light. However in the JET instrument an additional source is found in the laser beam penetration into the Torus Hall which is only a few centimeters away from the TS light collection penetration. Stray light from this penetration is in the field of view of the light collection optics in the Torus Hall, which is collecting it and sending it back into the spectrometer. Multiple reflections between the vacuum windows and collection mirrors produce a time delay so that certain reflections are mixing up with the useful TS signal. The required stray light rejection of the spectrometer (~$10^{-4}$ – $10^{-6}$) is mainly dictated by the intensity of the spurious reflections which must be suppressed.

## 3. JET LIDAR spectrometer

### 3.1 General scheme

The current spectrometer lay-out is slightly different from what is described in [2] and a schematic is show on figure 3.

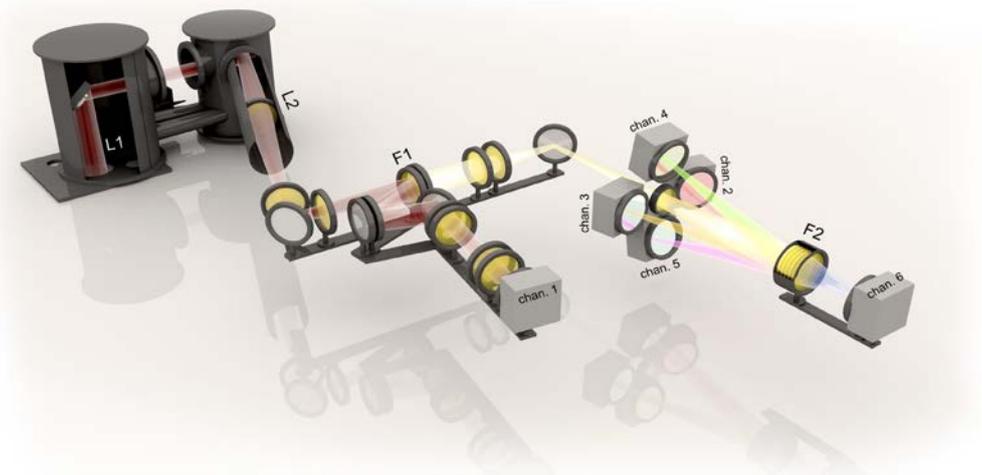

Figure 3: The JET LIDAR spectrometer

A lens L1 at the end of the neutron shield penetration (below the left vertical metal tube on figure 3) is imaging the 6 horizontal mirrors in the Torus Hall to a field lens marked as L2, which in turn is being relayed to the first shortpass interference filter F1 with a cut-off wavelength just below the ruby laser line. It therefore reflects the ruby light and all TS light of higher wavelength into the 1$^{st}$ detector, which has 2 stray light filters providing ~$10^{-6}$ rejection. The lower wavelength branch is relayed to the filter stack F2, which consists of two wedged substrates coated on either side as shortpass filters with different cut-offs (described in [2] but they were only installed in the spectrometer in 2013). Light is reflected into channels 2-5 depending on wavelength, and the shortest wavelength band (<500nm) is going through to spectral channel 6.

The physical size of the spectrometer is around 4m x 2m x 1m excluding the 2 vertical metal tubes and the first 2 lenses. The diameter of all major optical components is 15cm with about 1m distance between the subsequent imaging planes gives a theoretical maximum etendue of the spectrometer around 500 sr*mm$^2$, but in reality only a small fraction of that is being used. The biological shield penetration is narrower than it was planned when the spectrometer was designed, therefore the maximum light spot size at the pupil is almost twice as small as the lens size, which decreases the actual etendue by about factor of 4 to ~125sr*mm$^2$. Even that number is overestimated, since the image which is being relayed to all the interference filters and detectors is not a uniform circle (see figure 4) with considerable areas not illuminated (i.e. not used). Etendue of the actually collected light is about 40sr*mm$^2$, i.e. 10 times smaller than the spectrometer could have handled.

Despite the apparent inefficiency in using the spectrometers' optical power, relaying the 6-spot image to the filters and detectors serves an important purpose: shortpass filters with such a spectral range of transmission/rejection require many dielectric layers to be coated and making such a coating totally uniform over a large area is difficult if not impossible. The spectral transmission of each filter used on JET *is indeed different* at different areas, therefore the actual average transmission depends on the illumination of its surface. Since an image of the tokamak windows and not the plasma itself is being relayed – the scattered light footprint on each surface remains the same for each radial location along the collection path, and hence the spectral response of the spectrometer remains homogeneous and can be easily calibrated.

Another important aspect of such spectrometer design is stray light rejection. The large backwall stray light spike is at least $10^9$ times stronger than the main signal (as measured on JET by stacking neutral density filters) and to completely neglect it one would require ~$10^{-11}$ rejection. If only one-millionth part of it by diffuse scattering will come to a detector few nanoseconds earlier via some short-cut route, and mixes up with TS light, we would need to add an extra $10^{-5}$ ruby light rejection to that channel. For a high etendue spectrometer this is another expensive component to add to each spectral channel and leads to extra light losses. JET LIDAR spectrometer is designed in a way which excludes any possible shortcut routes for the backwall stray light flash. Only two spectral channels (1+2) are using extra stray light rejection filters, others can function unprotected.

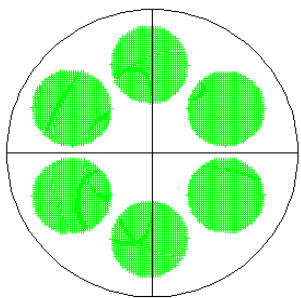

Figure 4: Image of the JET machine windows on the detector surface (d=11mm), as in the ray-tracing model

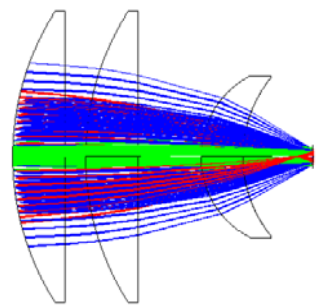

Figure 5: Ray-tracing model of the image compression on to the d=11mm detector surface. Light of different colours is coming from different positions in plasma.

## 3.2 Detectors

Fast GaAsP detectors (Hamamatsu 3809U-73A) with a sensitive area of d~11mm were previously used for the edge LIDAR on JET [5,6]. Operation of that diagnostic was suspended in 2009 and therefore these detectors become available for the core LIDAR. The main challenge for replacing the MA-2 detectors by the new ones was to adapt the image size to a smaller area (from 18mm to 11mm), which required going from f/1.2 imaging to ~f/0.75. This was achieved by using a strong meniscus lens in addition to the existing doublet (see figure 5). The lens had to be made of a glass with high refractive index: N-SF6 was chosen (n~1.8). Note that the glass spectral transmission is limited to $\lambda>400nm$.

## 3.3 Notes on calibration

Spectral calibration of the instrument is described in [2] and hasn't changed much since then. Additional procedures were introduced to take into account non-uniformity of the interference filters and detectors response. Each of the six apertures of the spectrometer is now illuminated independently and the final relative sensitivity of spectral channels is calculated as a combination of them. In figure 6 one can see the results of the 6 calibration curves and observe the differences between them.

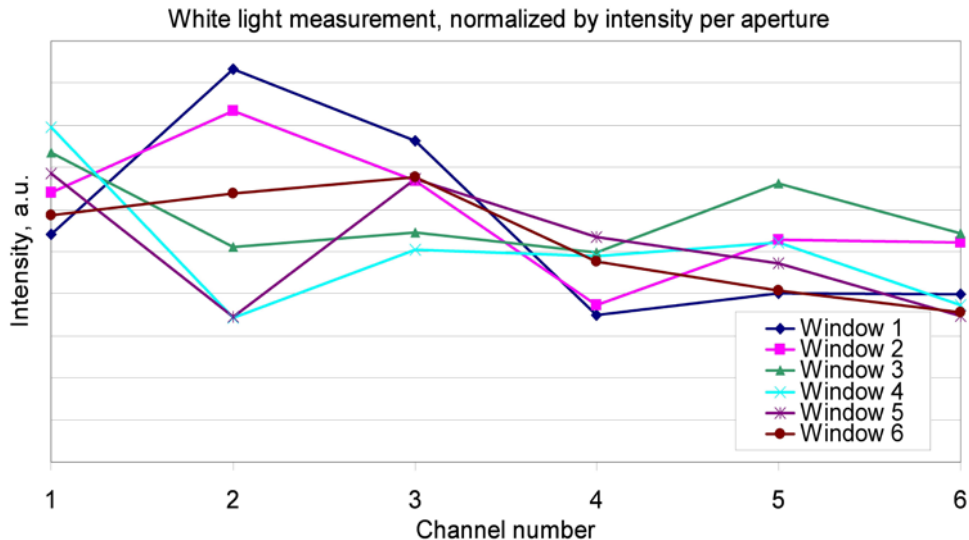

Figure 6: Spectrometer response to a calibrated white light source at each of the 6 collection windows

Calibration of the density profile shape, i.e. the light throughput at different radial plasma positions, is done using Raman scattering from nitrogen at ~400mbar pressure. A highly doped Ruby crystal [7] is used to suppress the parasitic laser light, while the interference filters used during normal plasma measurements are removed. The Ruby filter is composed of 64 small cubes 5x5x15mm glued to each other, therefore only the light which incidents almost normally to its surface is propagating through undisturbed. This filter also has a relatively small total aperture (d~40mm), which is not enough to cover the whole input of the spectrometer. The only way to effectively use it for calibration is to install it just after the L2 lens (see figure 3) where the variation in angle is the lowest and the image is composed of 6 independent light spots about ~35mm diameter, which can be covered by the ruby filter separately. That way the Raman calibration has to be done 6 times and the final result will be a combination of them.

Only the shape of the measured density profile is calibrated via Raman scattering, the absolute value of the measured density is calculated by cross-calibration with far-infrared interferometer measurements. This is done once, at the beginning of each experimental campaign, and remains accurate for months within a few percent.

Measuring Raman scattered light is also possible in a normal setup, with only interference filters rejecting the laser stray light, instead of the above mentioned ruby filter. A comparison between the two types of calibration is shown on figure 7, as one can see the results are different. The measurement with the interference filters is wrong, this is caused by sensitivity of the interference filters' cut-off wavelength to the angle of incidence, which is different for scattered light coming from different radial positions inside JET. This means that it is not possible to accurately perform the Raman calibration in the current JET LIDAR setup, without the ruby filter. To minimize the error one would need to reduce the incidence angle variations on the filters thus increasing the filter diameter even further, implications of that for the ITER LIDAR design are discussed in [8]. Note that the property of passive absorption of the laser wavelength by the laser crystal material is unique to 3-level systems such as Ruby, if Nd:YAG is used as a laser source then the method implemented on JET for the stray light suppression will not work.

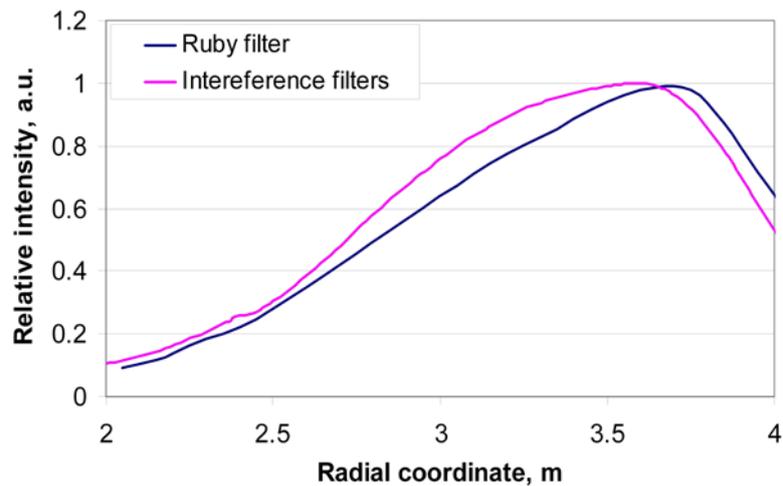

Figure 7: Results of the vignetting calibration, using ruby filter and dielectric interference filters for rejection of the parasitic light, both curves have been normalized.

## 4. Summary of recent modifications to the diagnostic.

In order to achieve better measurement accuracy, a number of changes were made to the JET LIDAR diagnostic. A review of the standard calibration scheme [9,10] has led to the introduction of new calibration procedures, e.g. calibration of the 6 apertures individually. The vacuum windows spectral transmission has been measured regularly since 2005 (figure 8). Thin coatings on the windows have a chromatic effect (transmission at blue wavelengths is worse than at red) which produces ~5% bias of the measured Te if not taken into account.

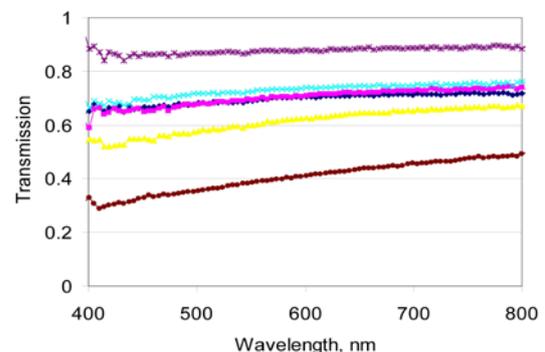

Figure 8: spectral transmission of 6 LIDAR light collection window

Lens L2 (figure 3) was initially an achromat used to operate a single-point Thomson Scattering diagnostic in parallel with LIDAR while the latter was being commissioned. This achromat consisted of 2 lenses made of BK7 and SF2 glass bonded to each other, with no AR coating on the outer surfaces. Since the single point TS system was decommissioned long ago, the achromat has recently been replaced by a simple BK7 plano-convex lens with a similar focal distance and AR coated for 400-800nm, which recovered about 8% signal in channels 1-5 and about twice as much in channel 6 due to poorer transmission of SF2 in that wavelength region.

A new collection optics alignment scheme was developed and used to improve the light collection efficiency. Implementation of the scheme during the 2009-2011 JET shutdown has shown that virtually no adjustment was required for 5 out of 6 collection sub-systems, but the last one was completely lost for unknown reasons and for an unknown period of time (first noticed in 2007). Alignment was restored and therefore $1/6^{th}$ of the total collected light was recovered.

The old four plane plate filters at stack F2 (figure 3) [2] were replaced by wedges coated on both sides. The plane filters did not have an AR coating on their back side which created unnecessary light losses especially for high temperature channels, for example to reach spectral channel 5, scattered light had to pass through the filters five times: 2-3-4-3-2.

GaAsP detectors were installed as described in section 3.2, with improved effective QE especially at 600-800nm where previous MA-2 detectors are very inefficient.

A large aperture, high transmission (~85-90%) nanowire polarizer was installed in the system - in the past no polarization filtering was used. A factor of 2 reduction of plasma background light had a strong positive impact, especially for high-Te measurements in H-mode plasmas where beryllium spectral lines are overwhelming spectral channels 5 and 6 (400-550nm). This problem was significantly enhanced after installation of the beryllium/tungsten ITER-like metal wall on JET [11], due to the increased Be concentration in the plasma and possibly even more enhanced by background light reflections from the metal wall inside the new vessel.

## 5. Overview of the produced measurements after the upgrade.

In figures 9-11 one can find examples of the most recent LIDAR measurements in different plasma conditions, from very small to very large stored energies, together with estimated relative errors. Displayed uncertainties of the measurements are assuming only statistical errors and are calculated with [12], using real signal to noise ratios in all spectral channels. Statistical error bars are designating a +/- 33% confidence interval. One can see that the quality of the measurement is varying at different radial coordinates, since light collection efficiency is changing as a function of radius (figure 2) and the accuracy of measurement at lower temperatures (<1keV) is reduced due to the stray light suppression requirements in the spectrometer design (scattered light near 694nm is lost) and spectral channels near the laser line being too wide because of the interference filters design.

In plasmas with low $n_e$ and $T_e$ the errors in temperature are still below 10% and below 5% in density for most of the plasma and become even better at R=3.0-3.5m where the diagnostic is most efficient (figure 9).

In typical plasmas of interest (figures 10 and 11), errors are going below 5% and 3% respectively for $T_e$ and $n_e$, which is far better than requested for the ITER LIDAR prototype (10% in Te and 5% density for $n_e>3*10^{19}$). Note that despite the more than a factor of two density increase between cases 2 and 3 (figures 10 and 11 respectively) the quality of measurement is almost unchanged, this is due to the increase of plasma background light which is growing together with density.

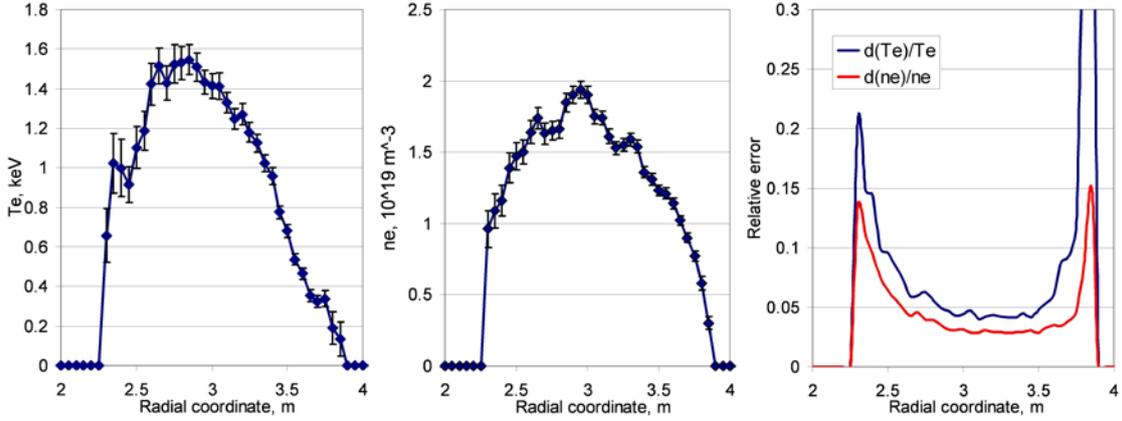

Figure 9: Example of LIDAR measurements, Ohmic low current plasma
#84909, t=55.37s, Ip=1.5MA, most unfavourable plasma for LIDAR

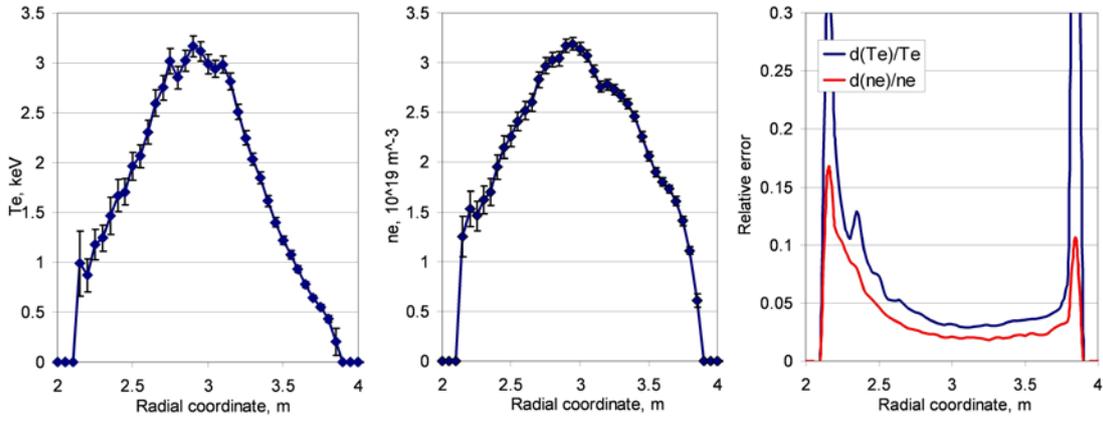

Figure 10: Example of LIDAR measurements, mild additional heating
#84710, t=59.63s P(NBI)=3MW, Ip=2.0MA

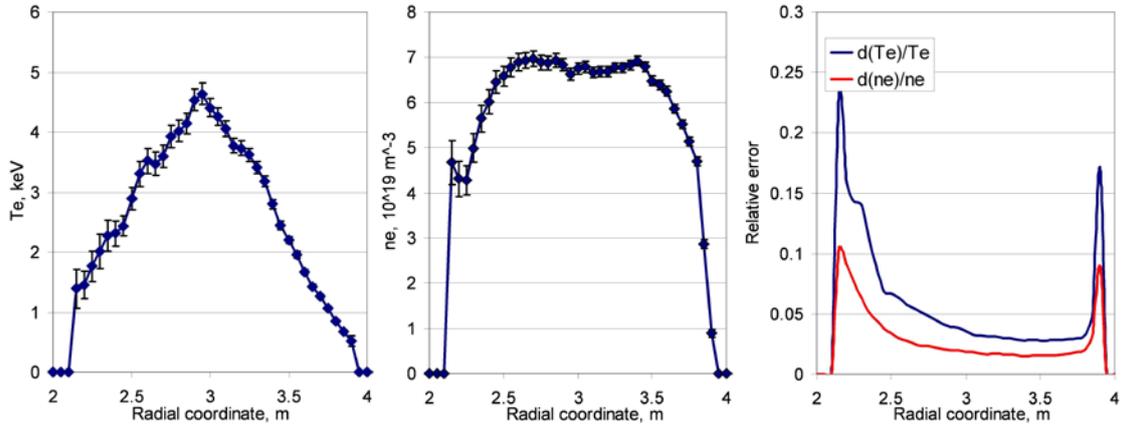

Figure 11: Example of LIDAR measurements, strong heating
#84779, t=52.87s, P(NBI)=20MW, Ip=3.0MA, solid baseline H-mode

In table 1 a summary of the current and previous status of the measurements is shown. Note that the indicated measurement accuracy only includes statistical errors (signal to noise ratio) and does not take into account improvements in calibration.

|  | Previous | Current |
| --- | --- | --- |
| Repetition rate | 4 Hz | 4 Hz |
| Measurements area | -0.75<r/a<0.75 | -0.4<r/a<0.95 |
| Spatial resolution | 12cm | 12cm |
| Statistical error in Te(core) for ohmic plasma | >20% | <5% |
| Statistical error in Ne(core) for ohmic plasma | >10% | <3% |
| Statistical error in Te(core) for heated plasma | ~10% | ~3% |
| Statistical error in Ne(core) for heated plasma | ~5% | ~1.5% |

Table 1: Status of measurements before and after the refurbishment

The spatial resolution remains the same and is currently limited by the slowest component – relatively old digitizers. A faster DAQ system is available on JET which would immediately allow us to achieve a resolution of ~7cm, but the implementation of the new system is currently pending a decision on the overvoltage protection.

Electron temperature on JET can be measured with three independent diagnostics: Electron Cyclotron Emission (ECE) [13], core LIDAR and conventional High Resolution Thomson Scattering (HRTS)[14]. The core density profile is in turn measured by LIDAR, HRTS and a microwave reflectometer [15]. The absolute calibration of density is somewhat tricky, the absolute value for both Thomson scattering diagnostics is cross-calibrated versus line integrated measurements of interferometry, which is considered to be accurate and reliable. The density profile shape of LIDAR is calibrated via Raman scattering as described above, HRTS density shape is in turn cross-calibrated to LIDAR in the core and reflectometer at the edge. Reflectometry is self-sufficient in terms of absolute density, but its profile position is dependant on magnetic field measurements and prone to be inaccurate, therefore it has to be adjusted to the best match with interferometer edge measurements.

In Figure 12 one can find an example of all the above mentioned diagnostics plotted on the same equatorial plane coordinate axis – there is a very good agreement between all of them. Discrepancies between the diagnostics do occur as a result of a number of physical and technical reasons, which are outside of the scope of this paper, although the variations as observed during the experiments performed in 2011-2012 typically are within a 5% margin.

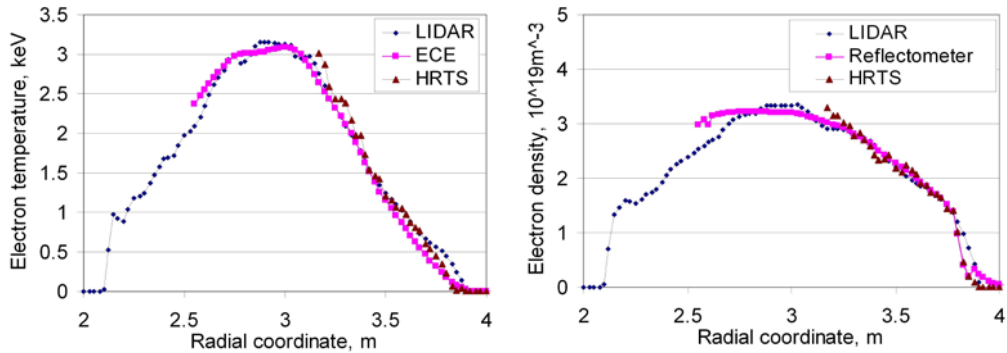

Figure 12: Comparison of various temperature and density measurements available on JET, #84710, t=52.80-52.90s.

## 6. Conclusions

The JET Core LIDAR diagnostic produced its first results more than 25 years ago and is still operational. Despite a good record on JET, it remains unique in the fusion world as LIDAR requires a minimum size device in order to obtain a spatial resolution of <10% of the minor radius. The LIDAR concept becomes more attractive for bigger fusion devices with restricted access due to neutron radiation/activation, in that case the simplicity of light collection optics and single port access to burning plasma, offered by LIDAR, becomes so compelling that it may outweigh the difficulties associated with laser and spectrometer/detectors design and development.

JET experience has shown that once the diagnostic is built, it is fairly easy to maintain and is capable of reliably producing data for many years. After the series of refurbishments done in 2011-2013, the diagnostic is producing measurements of the highest quality ever, well surpassing the accuracy requested for the ITER Core Thomson scattering. The diagnostic is expected to stay operational in future JET experiments and produce measurements in the forth-coming full scale DT campaign.

## Acknowledgments

This work, part-funded by the European Communities under the contract of Association between EURATOM/CCFE, was carried out within the framework of the European Fusion Development Agreement. For further information on the contents of this paper please contact publications-officer@jet.efda.org.* The views and opinions expressed herein do not necessarily reflect those of the European Commission. This work was also part-funded by the RCUK Energy Programme under grant EP/I501045